\documentclass{aastex63}

\usepackage{amsmath}

\newcommand{\elsasser}{Els\"asser\ }
\newcommand{\alfven}{Alfv\'en\ }

\received{}
\revised{}
\accepted{}
\submitjournal{ApJ}

\shorttitle{Non-linear damping of standing kink waves}
\shortauthors{Van Doorsselaere et al.}
\begin{document}

\title{Non-linear damping of standing kink waves computed with Els\"asser variables}

\correspondingauthor{Tom Van Doorsselaere}
\email{tom.vandoorsselaere@kuleuven.be}

\author[0000-0001-9628-4113]{Tom Van Doorsselaere}
\affiliation{Centre for mathematical Plasma Astrophysics, Department of Mathematics, KU~Leuven, Celestijnenlaan 200B, B-3001 Leuven, Belgium}

\author{Marcel Goossens}
\affiliation{Centre for mathematical Plasma Astrophysics, Department of Mathematics, KU~Leuven, Celestijnenlaan 200B, B-3001 Leuven, Belgium}

\author{Norbert Magyar}
\affiliation{Centre for Fusion, Space and Astrophysics, Department of Physics, University of Warwick, CV4 7AL, Coventry, UK}

\author{Michael S. Ruderman}
\affiliation{School of Mathematics and Statistics (SoMaS), The University of Sheffield, Hicks Building, Hounsfield Road, Sheffield, S3 7RH, UK}
\affiliation{Space Research Institute (IKI) Russian Academy of Sciences, Moscow, Russia}

\author{Rajab Ismayilli}
\affiliation{Centre for mathematical Plasma Astrophysics, Department of Mathematics, KU~Leuven, Celestijnenlaan 200B, B-3001 Leuven, Belgium}



\begin{abstract}
	In a previous paper, we computed the energy density and the non-linear energy cascade rate for transverse kink waves using \elsasser variables. In this paper, we focus on the standing kink waves, which are impulsively excited in coronal loops by external perturbations. We present an analytical calculation to compute the damping time due to the non-linear development of the Kelvin-Helmholtz instability. The main result is that the damping time is inversely proportional to the oscillation amplitude. \\
	We compare the damping times from our formula with the results of numerical simulations and observations. In both cases we find a reasonably good match. The comparison with the simulations show that the non-linear damping dominates in the high amplitude regime, while the low amplitude regime shows damping by resonant absorption. In the comparison with the observations, we find a power law inversely proportional to the amplitude $\eta^{-1}$ as an outer envelope for our Monte Carlo data points. 
\end{abstract}

\keywords{MHD -- solar corona -- MHD waves -- turbulence}


\section{Introduction} \label{sec:intro}
Magnetohydrodynamic (MHD) waves are ubiquitously present in the solar atmosphere and magnetosphere \citep{nakariakov2016}. A particular kind of MHD waves are the transverse kink waves in coronal loops. They have been observed since more than two decades \citep{nakariakov1999,schrijver1999,aschwanden1999}. These transverse waves are attractive to observe and model because of two reasons: (1) they can be used for coronal seismology with the aim of estimating physical parameters in coronal loops \citep[e.g.][see also the review by \citeauthor{nakariakov2020} \citeyear{nakariakov2020}]{pascoe2014,magyar2018,pascoe2018}, and (2) they could play a role in heating the corona \citep[e.g.][see also the reviews by \citeauthor{arregui2015} \citeyear{arregui2015}; \citeauthor{vd2020ssrv} \citeyear{vd2020ssrv}]{demoortel2012,terradas2018b,karampelas2019b,pagano2019,hillier2020}. \par

In this work, we focus on standing transverse waves in coronal loops. Nowadays, we understand that these come in two flavours. On the one hand, there are the decayless oscillations, as discovered by \citet{wang2011,tian2012,nistico2013}. These have no apparent external excitation source and many researchers presume they are driven \citep[at the footpoints?,][]{karampelas2017,karampelas2018,afanasyev2020} to maintain the quasi-steady amplitude. It was shown observationally that their period scales with the loop length \citep{anfinogentov2015}, conclusively showing that these are also standing waves. On the other hand, there are the transverse waves in coronal loops that are impulsively excited by a flare \citep{aschwanden2011} or low coronal eruption \citep{zimovets2015}. These oscillations show an initial displacement from their equilibrium position by the external exciter, after which the coronal loop apparently freely oscillates with its natural frequency showing a strong damping.  \\
In this work, we will consider the damping of the impulsively excited transverse waves, and leave the decayless waves aside for now. \par

For the early observations of impulsively excited standing kink waves, their strong damping was interpreted as damping in terms of resonant absorption \citep{ruderman2002,goossens2002}. This phenomenon is an ideal damping mechanism \citep{terradas2006} that converts over time the observed, large-scale, coherent transverse motions to localised, incoherent motions around the resonant layer \citep{soler2013,goossens2014}, which are hard to observe, resulting in an apparent damping. Resonant absorption is a damping that works in linearised MHD, leading to damping with exponential behaviour that is independent of the amplitude of the wave \citep[see e.g.][]{goossens1992}. \\
In recent years, it was found that the situation is more complicated than simple exponential damping. In simulations, it was shown that the damping starts with a Gaussian phase first \citep{pascoe2012}, which was later confirmed with analytical calculations \citep{hood2013,ruderman2013}. Gaussian damping was also found in observations \citep{pascoe2016} and used for Bayesian seismology \citep{arregui2013b, pascoe2017}. Nowadays, the general damping profile has been characterised with numerical simulations \citep{pascoe2019}. \\
Despite all this progress, the damping is independent of the amplitude, and non-linear effects are not considered. 

It has been speculated already in the 80s that transverse kink waves are susceptible to the Kelvin-Helmholtz instability \citep[KHI, ][]{hollweg1988}. More recently, it has been confirmed numerically that standing transverse oscillation lead to the non-linear development of the KHI \citep{terradas2008b}, resulting in the formation of so-called transverse wave induced Kelvin-Helmholtz rolls \citep[or TWIKH rolls for short,][]{antolin2014,antolin2016,vd2018}. \citet{magyar2016b} constructed similar numerical models and investigated in a parametric study the damping time as a function of the initial amplitude. They found that low amplitude oscillations indeed follow the damping by resonant absorption, but that at high amplitudes the non-linear damping takes over and a strong amplitude dependence was found. Physically this can be understood as the cascade of wave energy at large scales to the smaller scale TWIKH rolls, once again leading to apparent damping. \\
Also in the meta-analysis of \citet{goddard2016}, which considered 25 observed transverse loop oscillation events, it was found that the damping time depends on the oscillation amplitude. Their initial estimate was that the damping time $\tau$ (normalised to the period $P$) would have an upper limit of $\eta^{-1/2}$, with $\eta$ the displacement amplitude. The study was later extended by \citet{nechaeva2019}, who collected data over the entire solar cycle, resulting in more than 200 cases. They found once again a strong observational dependence of the damping on the amplitude, and the fitted an upper limit for the damping time of $\tau/P\sim \eta^{-0.68}$. \par

The non-linear aspects of kink modes have been considered in the past, but they were mostly focused on the calculation of eigenfunction modifications and period changes \citep{ruderman2014}. For example, \citet{ruderman2017} found that the non-linear effect of kink waves is to generate fluting modes with azimuthal wave number $m=2$. This was later confirmed numerically by \citet{terradas2018}. Moreover, some models studied the KHI instability criterion in oscillating loops \citep{hillier2019,barbulescu2019}.\\
Previously, the evolution of propagating waves and their energy was modeled by \citet{ruderman2010}. They found extra damping compared to the linear regime, as a consequence of the small scale development by the non-linear effects. Motivated by the discovery of the uniturbulent regime \citep{magyar2017}, the non-linear evolution of (propagating and standing) kink waves was recently revisited by \citet{vd2020}. They used a formulation in terms of \elsasser variables to estimate the wave energy density and turbulent energy cascade rate \citep[similar to common practices in solar wind modelling,][]{bruno2013,vanderholst2014}. They found a damping time for the propagating wave, which was inversely proportional to the amplitude, i.e. $\eta^{-1}$.\\
Here we investigate how the results in \citet{vd2020} extend to standing waves. We compare our results to the numerical parametric study of \citet{magyar2016b} and the observational results of \citet{nechaeva2019}.

\section{Earlier results}
In \citet{vd2020}, the simplest model for a magnetic field-aligned, overdense cylinder was considered (uniform magnetic field $B$, internal/external density $\rho_\mathrm{i/e}$). There the transverse kink waves are mathematically described by Bessel functions, leading to the usual dispersion relation \citep{zaitsev1975,wentzel1979,edwin1983}. \\
\citet{vd2020} found the expressions for the energy density $w$ in standing or propagating kink waves in the thin-tube limit $\delta=k_zR\ll 1$, where $k_z$ is the longitudinal wave number and $R$ is the radius of the loop. In that approximation, the radial Bessel eigenfunctions ${\cal R}(r)$ for the total pressure  perturbation $P'$ reduce to 
\begin{equation}
	\lim_{\delta\to 0} {\cal R}(r)=\begin{cases} A\frac{r}{R} & \mbox{for } r\leq R\\ A\frac{R}{r} & \mbox{for } r>R\end{cases},
\end{equation}
with oscillation amplitude $A$. They found expressions for the \elsasser variables \begin{equation} \vec{z}^\pm=\vec{v}\pm \frac{\vec{b}}{\mu \rho}\label{eq:linearelsasser} \end{equation} for the perturbations, where $\vec{v}$ and $\vec{b}$ are the velocity and magnetic field perturbation and the + (-) represent the downward (upward) travelling Alfv\'en wave in a uniform medium. Their expressions for the associated energy density to these \elsasser variables for kink waves are given in their equations 52 \& 53 as
\begin{align}
	w^\pm_\mathrm{i} &= \frac{1}{4}\frac{1}{\rho_\mathrm{i} (\omega^2-\omega_\mathrm{Ai}^2)^2}\frac{A^2}{R^2} \begin{cases}  (\omega \cos{k_z z} \sin{\omega t}\pm \omega_\mathrm{Ai} \sin{k_z z} \cos{\omega t})^2 & \mbox{(standing)} \\ (\omega \mp \omega_\mathrm{Ai})^2 \sin^2{(k_z z-\omega t)} & \mbox{(propagating)} \end{cases},\\
	w^\pm_\mathrm{e} &= \frac{1}{4}\frac{1}{\rho_\mathrm{e} (\omega^2-\omega_\mathrm{Ae}^2)^2}\frac{A^2R^2}{r^4} \begin{cases}  (\omega \cos{k_z z} \sin{\omega t}\pm \omega_\mathrm{Ae} \sin{k_z z} \cos{\omega t})^2 & \mbox{(standing)} \\ (\omega \mp \omega_\mathrm{Ae})^2 \sin^2{(k_z z-\omega t)} & \mbox{(propagating)}\end{cases}. \label{eq:wintwext}
\end{align}
In these expressions,  subscripts $i/e$ correspond to the internal and exterior region, the oscillation frequency is $\omega$, the \alfven frequency $\omega_\mathrm{A}$ is related to the \alfven speed $V_\mathrm{A}$ through $\omega_\mathrm{A}=k_zV_\mathrm{A}$. In these formulae, the top line corresponds to the standing wave, while the bottom line corresponds to propagating waves, as indicated.

As explained in \citet{vd2020}, the energy cascade rate is given by \begin{equation}
	\epsilon^{\mp}=\frac{\rho}{2}\vec{z}^{\,\mp}\cdot (\vec{z}^{\,\pm}\cdot \nabla \vec{z}^{\,\mp})=\vec{z}^{\,\pm}\cdot \nabla w^\mp, \qquad \mbox{with } w^\mp = \frac{\rho (\vec{z}^{\,\mp})^2}{4}.\label{eq:cascaderate}
\end{equation}
They found that the dominant contribution to the energy cascade rate is due to
\begin{equation} 
	\epsilon^\mp = z^\pm_{r\mathrm{e}} \frac{\partial}{\partial r} w^\mp_e,
\end{equation}
because these are the only terms that have a contribution at $\delta^3=k_z^3R^3$. In this expression, $z^\pm_{r\mathrm{e}}$ is the radial component of the \elsasser variable in the exterior plasma. \citet{vd2020} computed 
\begin{equation}
	\epsilon^\mp = \frac{A^3R^3}{r^7} \frac{1}{\rho_\mathrm{e}^2(\omega^2-\omega_\mathrm{Ae}^2)^3} \cos{\varphi} \begin{cases}  (-\omega \cos{k_z z} \sin{\omega t}\mp \omega_\mathrm{Ae} \sin{k_z z} \cos{\omega t})(\omega \cos{k_z z} \sin{\omega t}\mp \omega_\mathrm{Ae} \sin{k_z z} \cos{\omega t})^2 & \mbox{(st.)} \\ (\omega \mp \omega_\mathrm{Ae})(\omega \pm \omega_\mathrm{Ae})^2 \sin^3{(k_z z-\omega t)}& \mbox{(pr.)}\end{cases}
\end{equation}
in their Eq.~55.
To obtain the damping rate for propagating waves under uniturbulence \citep{magyar2017}, \citet{vd2020} divided the average energy density by the average energy cascade rate:
\begin{equation}
	\tau=\frac{\langle w^++w^-\rangle}{\langle \epsilon^++\epsilon^-\rangle}
\end{equation}
where the averaging was over the cross-section and period:
\begin{equation}
\langle \epsilon \rangle = \int_0^\infty r dr \left( \int_0^{2\pi} d\varphi \frac{\omega}{2\pi}\int_0^{2\pi/\omega}dt \ \epsilon^2 \right)^{1/2} \label{eq:avenergy}
\end{equation}
Intermediate results by \citet{vd2020} are (their Eq.~56):
\begin{equation}
	\epsilon = \epsilon^++\epsilon^-= \frac{A^3R^3}{r^7} \frac{1}{\rho_\mathrm{e}^2(\omega^2-\omega_\mathrm{Ae}^2)^3} \cos{\varphi} \begin{cases}  -2\omega \cos{k_z z} \sin{\omega t}(\omega^2 \cos^2{k_z z} \sin^2{\omega t}- \omega_\mathrm{Ae}^2 \sin^2{k_z z} \cos^2{\omega t}) & \mbox{(st.)}\\ 2\omega(\omega^2 - \omega_\mathrm{Ae}^2) \sin^3{(k_z z-\omega t)} &\mbox{(pr.)}\end{cases}
\end{equation}
and (their Eqs.~57-58)
\begin{align}
	\langle \epsilon \rangle &= \frac{A^3R^3}{\rho_\mathrm{e}^2(\omega^2-\omega_\mathrm{Ae}^2)^3} \frac{1}{5R^5} \left(\sqrt{\pi}\right) \begin{cases}  \frac{1}{2}\vert \omega \cos{k_z z}\vert \sqrt{ 4\omega^4 \cos^4{k_z z}+(\omega^2 \cos^2{k_z z}- \omega_\mathrm{Ae}^2 \sin^2{k_z z})^2} & \mbox{(st.)}\\ 2\vert\omega(\omega^2 - \omega_\mathrm{Ae}^2)\vert \sqrt{\frac{5}{16}} &\mbox{(pr.)}\end{cases}\\
	& = V^3 \frac{\sqrt{\pi} R}{10} \frac{\rho_\mathrm{e}}{\omega^3} \begin{cases}  \vert \omega \cos{k_z z}\vert \sqrt{ 4\omega^4 \cos^4{k_z z}+(\omega^2 \cos^2{k_z z}- \omega_\mathrm{Ae}^2 \sin^2{k_z z})^2} & \mbox{(st.)}\\ \sqrt{5}\vert\omega(\omega^2 - \omega_\mathrm{Ae}^2)\vert &\mbox{(pr.)}\end{cases} \label{eq:energycascade}
\end{align}
The quantity $V$ denotes the velocity amplitude in the interior region of the loop. \\ We can compute from the previous expressions that 
\begin{align}
	\langle \epsilon^\mp \rangle &= \frac{A^3R^3}{\rho_\mathrm{e}^2(\omega^2-\omega_\mathrm{Ae}^2)^3} \frac{1}{5R^5} \left(\sqrt{\pi}\right) \nonumber \\ & \begin{cases}  \frac{1}{4}\sqrt{(\omega^2 \cos^2{k_z z}+\omega_\mathrm{Ae}^2 \sin^2{k_z z})( 5(\omega^2 \cos^2{k_z z}+ \omega_\mathrm{Ae}^2 \sin^2{k_z z})^2-16 \omega^2\omega_\mathrm{Ae}^2 \cos^2{k_z z} \sin^2{k_z z})} & \mbox{(st.)}\\ \sqrt{\frac{5}{16}}\vert (\omega \mp \omega_\mathrm{Ae})(\omega \pm \omega_\mathrm{Ae})^2 \vert &\mbox{(pr.)}\end{cases}\\
	& = V^3 \frac{\sqrt{\pi} R}{20}  \frac{\rho_\mathrm{e}}{\omega^3} \begin{cases}  \sqrt{(\omega^2 \cos^2{k_z z}+\omega_\mathrm{Ae}^2 \sin^2{k_z z})( 5(\omega^2 \cos^2{k_z z}+ \omega_\mathrm{Ae}^2 \sin^2{k_z z})^2-16 \omega^2\omega_\mathrm{Ae}^2 \cos^2{k_z z} \sin^2{k_z z})} & \mbox{(st.)}\\ \sqrt{5}\vert \omega^2 - \omega_\mathrm{Ae}^2\vert \vert \omega \pm \omega_\mathrm{Ae}\vert &\mbox{(pr.)}\end{cases} \label{eq:energypm}
\end{align}
Curiously enough, the expression for the standing wave does not depend on the sign of the \elsasser variable! Probably this means that the damping for the $z^-$ is as strong as for $z^+$, because they are both part of the same standing wave.

\section{Non-linear damping of kink waves}\label{sec:damping}
In order to calculate the energy density and energy cascade rate of the standing kink waves, we extend expression \ref{eq:avenergy} to also average over the wavelength:
\begin{equation}
	\langle \langle \epsilon \rangle \rangle = \int_0^\infty r dr \left( \int_0^{2\pi} d\varphi \frac{k_z}{2\pi}\int_{-\pi/k_z}^{\pi/k_z}dz \frac{\omega}{2\pi}\int_0^{2\pi/\omega}dt \ \epsilon^2 \right)^{1/2}. \label{eq:avenergyz}
\end{equation}
We sum the contributions of both \elsasser components in each of the interior and exterior region:
\begin{equation}
	\langle \langle w_\mathrm{i/e}\rangle \rangle  = \langle \langle w^+_\mathrm{i/e}+w^-_\mathrm{i/e}\rangle \rangle
\end{equation}
and we obtain a wave energy density of 
\begin{equation}
	\langle \langle w \rangle \rangle = \langle \langle w_\mathrm{i}\rangle \rangle +\langle \langle w_\mathrm{e}\rangle \rangle =\pi R^2 \frac{\rho_\mathrm{i}+\rho_\mathrm{e}}{4}V^2, \label{eq:energy}
\end{equation}
which is half the energy of a propagating wave \citep[Eq. 59 in][]{vd2020} with the same amplitude. This can be understood intuitively, since a standing wave of the same amplitude as the propagating wave is the superposition of two propagating waves with half the amplitude. For the energy, it implies 2$\times (1/2)^2$, resulting in the factor 1/2. \par

To compute the energy cascade rate, we start from the intermediate result in Eq.~\ref{eq:energycascade}. It is mathematically arbitrary to first sum $\epsilon^\pm$ before averaging it (this operation is non-commutative). However, physically it is preferable to first compute the total energy cascade rate before averaging it over space and time. The reason for this is that the standing wave is operating as a whole: the kink wave contains both the $\vec{z}^\pm$ components simultaneously, and their combined damping is responsible for attenuating the wave. Thus, both quantities must both be summed first before averaging over space and time. Even though we think it is incorrect, we have included the alternative result is (with the averaging first, before the summing) in appendix~\ref{sec:appendix}.\\
Continuing from the summed energy cascade rates in Eq.~\ref{eq:cascaderate}, we have subsequently:
\begin{align}
	\langle\langle \epsilon \rangle\rangle 
	& = V^3 \frac{\sqrt{\pi} R}{10} \frac{\rho_\mathrm{e}}{\omega^3} \left(\frac{k_z}{2\pi}\int dz\  \omega^2 \cos^2{k_z z} (4\omega^4 \cos^4{k_z z}+(\omega^2 \cos^2{k_z z}- \omega_\mathrm{Ae}^2 \sin^2{k_z z})^2)\right)^{1/2}\\
	&= V^3 \frac{\sqrt{\pi} R}{10} \frac{\rho_\mathrm{e}}{\omega^3} \left(\frac{k_z}{2\pi}\int dz (5\omega^6 \cos^6{k_z z}-2\omega^4\omega_\mathrm{Ae}^2 \cos^4{k_z z} \sin^2{k_z z}+\omega^2\omega_\mathrm{Ae}^4 \cos^2{k_z z} \sin^4{k_z z})\right)^{1/2}\\
	&= V^3 \frac{\sqrt{\pi} R}{10} \frac{\rho_\mathrm{e}}{\omega^3} \frac{1}{4}\left(25\omega^6 - 2\omega^4\omega_\mathrm{Ae}^2+\omega^2\omega_\mathrm{Ae}^4\right)^{1/2}\\
	&= V^3 \frac{\sqrt{\pi} R}{10} \frac{\rho_\mathrm{e}}{4}\left(25 - 2\left(\frac{\omega_\mathrm{Ae}}{\omega}\right)^2+\left(\frac{\omega_\mathrm{Ae}}{\omega}\right)^4\right)^{1/2}.
\end{align}
We use the approximate expression for the kink frequency
\begin{equation} \omega^2=\frac{\rho_\mathrm{i}\omega_\mathrm{Ai}^2+\rho_\mathrm{e}\omega_\mathrm{Ae}^2}{\rho_\mathrm{i}+\rho_\mathrm{e}}=\omega_\mathrm{Ae}^2\frac{2}{1+\zeta},
\end{equation}
with $\zeta=\rho_\mathrm{i}/\rho_\mathrm{e}$. In this equation, the last equality is only valid if the magnetic field is uniform. We can then further simplify the energy cascade rate to
\begin{align}
	\langle\langle \epsilon \rangle\rangle  &= V^3 \frac{\sqrt{\pi} R}{10} \frac{\rho_\mathrm{e}}{4}\left(25 - 2\left(\frac{1+\zeta}{2}\right)+\left(\frac{1+\zeta}{2}\right)^2\right)^{1/2}\\
	&= V^3 \frac{\sqrt{\pi} R}{10} \frac{\rho_\mathrm{e}}{8}\sqrt{\zeta^2-2\zeta+97}. \label{eq:cascresult}
\end{align}
Somewhat surprisingly\footnote{Perhaps it is even more surprising that the largest prime number below 100 would occur in the mathematical description of a physical phenomenon!}, this expression does not tend to 0 as $\zeta \to 1$, in contrast to the damping of propagating waves by uniturbulence \citep[Eq.~58 of][]{vd2020}. This agrees with the findings of \citet{howson2019} that the non-linear damping by KHI also works in a coronal loop model with a uniform density but varying magnetic field.\par

\begin{figure}
	\plotone{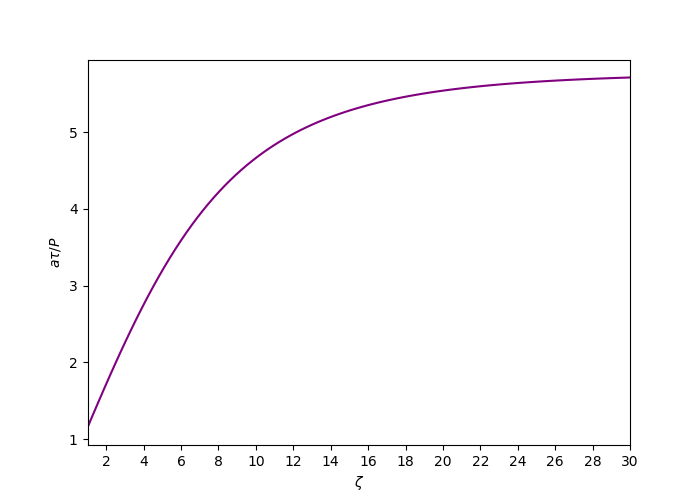}
	\caption{The damping time $a\tau /P$ due to non-linearity for a standing kink wave as a function of the density contrast $\zeta$, using Eq.~\ref{eq:tau2}.}
	\label{fig:tau2}
\end{figure}

Now we can find the expression for the damping time by dividing the energy density (Eq.~\ref{eq:energy}) by the energy cascade rate (Eq.~\ref{eq:cascresult}), using the same method as in \citet{vd2020}.
\begin{equation}
	\tau=\frac{\langle w\rangle}{\langle\langle \epsilon\rangle\rangle}=20\sqrt{\pi}\frac{R}{V}\frac{1+\zeta}{\sqrt{\zeta^2-2\zeta+97}}=20\sqrt{\pi}\frac{P}{2\pi a}\frac{1+\zeta}{\sqrt{\zeta^2-2\zeta+97}}.\label{eq:tau2}
\end{equation}
Here $a=\eta/R$ is the relative displacement amplitude, i.e. the ratio of the displacement $\eta$ compared to the loop radius $R$. The strongest damping occurs for $\zeta=1$ and is equal to $\tau/P=\frac{1}{a}\frac{20}{\sqrt{96\pi}}=1.15/a$. For an infinitely dense loop or vacuum exterior $\zeta\to \infty$, the damping saturates with a maximum of $\tau/P=10/a\sqrt{\pi}\approx 5.64/a$. The full graph of $\tau$ is shown in Fig.~\ref{fig:tau2}.

The damping time $\tau$ computes how fast the kink wave energy is cascaded to smaller scales. This formula considers the inertial regime of the turbulent cascade \citep[see Fig. 10 of][]{vd2020ssrv}, which is an ideal MHD process. In this turbulent cascade, the energy dissipation rate does not depend on the scale of the eddies, resulting in a power law behaviour (cfr. Kolmogorov or Iroshnikov-Kraichnan scaling). Once the turbulent eddies enter the dissipative range, the energy cascade rate will take a different form (which does depend on the dissipation coefficients) and also have a different power law slope. \\
The analogy with damping by resonant absorption is clear. The damping in resonant absorption is occuring also in ideal MHD. The details of viscous/resistive damping of the resonant Alfv\'en modes is different, and not needed to find the resonant damping rate (see Eq.~\ref{eq:ra}). In a sense, resonant absorption is a cascade to smaller radial wave numbers (albeit independent of the amplitude), reinforcing the analogy with the non-linear damping (Eq.~\ref{eq:tau2}) which represents a cascade in the radial and azimuthal wave number.

\section{Comparison to simulations}\label{sec:simulations}

%Using the parameters of \citet{antolin2014} with $\zeta=3$ and amplitude $a=0.44$, we find a damping time of $\tau/P\sim 10$, which is not too far off their simulated damping time of 

\begin{figure}
	\plotone{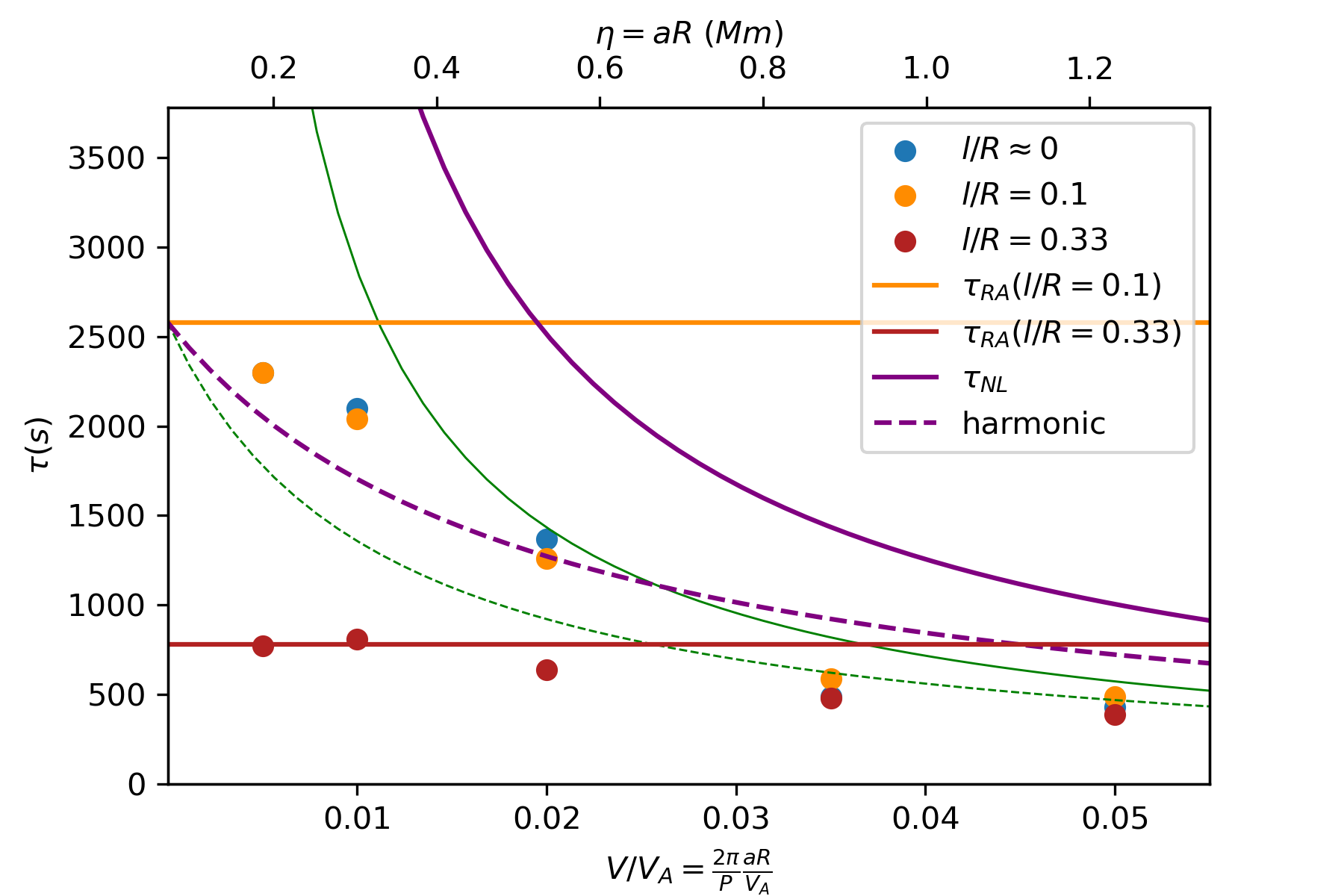}
	\caption{The damping times $\tau$ as a function of initial oscillation amplitude $V/V_\mathrm{Ai}$ (or equivalently $\eta$ in the top horizontal axis). The points correspond to the results of \citet{magyar2016b}, with the colour showing different $l/R$ as displayed in the legend. The full purple line are the results of Eq.~\ref{eq:tau2}. The horizontal red and orange lines are the damping times from resonant absorption (Eq.~\ref{eq:ra}), with the appropriate $l/R$ as indicated in the legend. The dashed purple line is the harmonic average of the resonant damping time for $l/R=0.1$ and the non-linear damping time. For completeness, the damping time with the alternative method in the appendix (Eq.~\ref{eq:tau1}) is shown with the full green line, and the harmonic average with the dashed green line.}
	\label{fig:magyar}
\end{figure}

Here we compare the results for the damping time (Eq.~\ref{eq:tau2}) to those  of \citet{magyar2016b}. In that work, 3D simulations were performed of standing kink waves, of which the amplitude and scale of the inhomogeneous layer $l/R$ was varied in a parameter study. The authors found that the damping is determined by resonant absorption for small amplitudes. In this regime, the damping is independent of the wave amplitude \citep{goossens1992}. However, in the high amplitude regime, they found that the damping was caused by the non-linear evolution of the kink mode, namely the formation of the KHI \citep{terradas2008b}. They considered a loop with an \alfven speed of $V_\mathrm{Ai}=0.6\mathrm{Mm/s}$, radius of $R=1.5\mathrm{Mm}$ and density contrast of $\zeta=5$.  They considered as amplitude parameter of the initial velocity perturbation the values $V=\lbrace 0.005,0.01,0.02,0.035,0.05\rbrace V_\mathrm{Ai}$, given in terms of the internal \alfven speed. Their results for the damping times as shown in their Fig.~6 are displayed as dots in Fig.~\ref{fig:magyar}, for three different values of the thickness of the inhomogeneous layer between the interior and exterior region $l/R=\lbrace 0,0.1,0.33\rbrace$.\par

On their data points, we overplot a few lines. The purple line corresponds to the damping time computed with the formula Eq.~\ref{eq:tau2}, which was derived for $l/R=0$. The orange and red horizontal lines are the expected damping times from the thin-tube, thin-boundary limit in resonant absorption \citep{ruderman2002} using a sinusoidal density profile in the inhomogenous layer:
\begin{equation}
	\frac{\tau_\mathrm{RA}}{P}=\frac{2}{\pi} \frac{R}{l} \frac{\zeta+1}{\zeta-1}, \label{eq:ra}
\end{equation}
for the parameters used in the \citet{magyar2016b} simulations.\\
From this graph, it is can be seen that the purple line captures the behaviour of the damping quite well in the region of high amplitude, apart from a vertical offset. The latter offset could be explained by the shortcomings of the present analytical model that was derived under the assumption that $l/R=0$. Another transition region from the interior to the exterior of the loop could easily alter the constant prefactor ($20\sqrt{\pi}$) of the non-linear damping, as it also does for resonant absorption \citep{soler2013}.\\ 
Another shortcoming is that the non-linear damping does not capture well the apparent saturation that occurs for small amplitudes. To address this, the dashed purple line shows the harmonic average of the non-linear damping and the damping by resonant absorption. In the dashed line, we have taken $1/\tau=1/\tau_\mathrm{RA}+1/\tau_\mathrm{NL}$, where we take $\tau_\mathrm{NL}$ from Eq.~\ref{eq:tau2}. A better correspondence with the numerical simulation points is indeed found. However, we still cannot match the observed inflection point in the simulation data. Consider the damping time's dependence on the wave amplitude $a$. For resonant absorption, we know $\tau_\mathrm{RA}\sim C$, while for the non-linear damping, we know $\tau_\mathrm{NL}\sim D/a$. Then, $\tau\sim CD/(aC+D)$. This is a rational function. Since both $C$ and $D$ are positive, it will have a pole at $a<0$. Beyond this, it is a monotonically decreasing section of the hyperbola. Thus, with the harmonic average it is always impossible to obtain an inflection point in this case.\\ 
In taking the harmonic average, we have assumed that the non-linear damping and resonant absorption operate independently from each other. However, it may be possible that the interaction between these damping mechanisms is more complicated (e.g. TWIKH rolls lengthen the resonant layer), and then a better fit could be obtained. 

For completeness, also the result obtained in appendix~\ref{sec:appendix} (in particular Eq.~\ref{eq:tau1}) is shown in Fig.~\ref{fig:magyar} with green, and its harmonic average is shown with dashed green lines.

\section{Matching with observations}\label{sec:monte}
\begin{figure}
	\plottwo{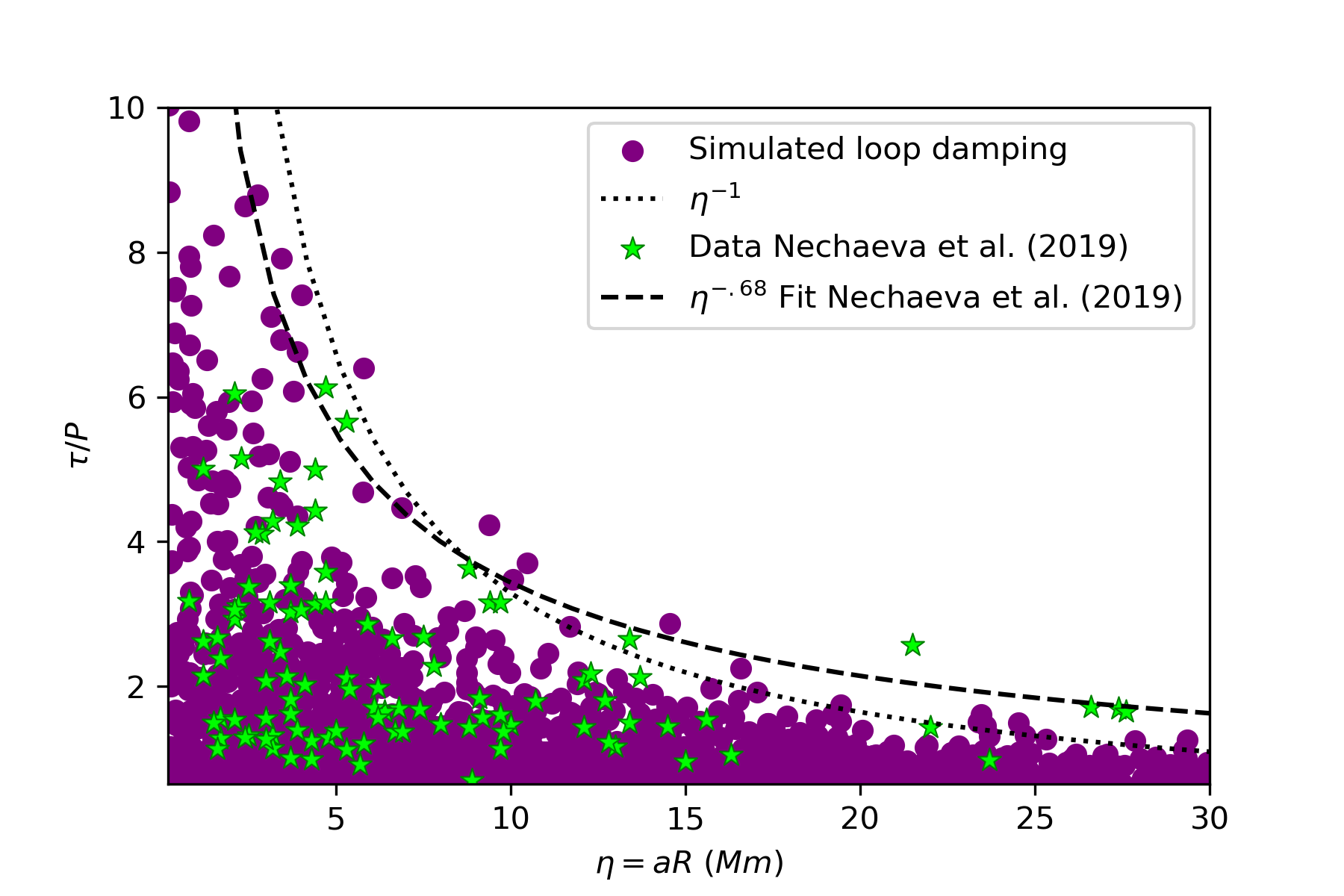}{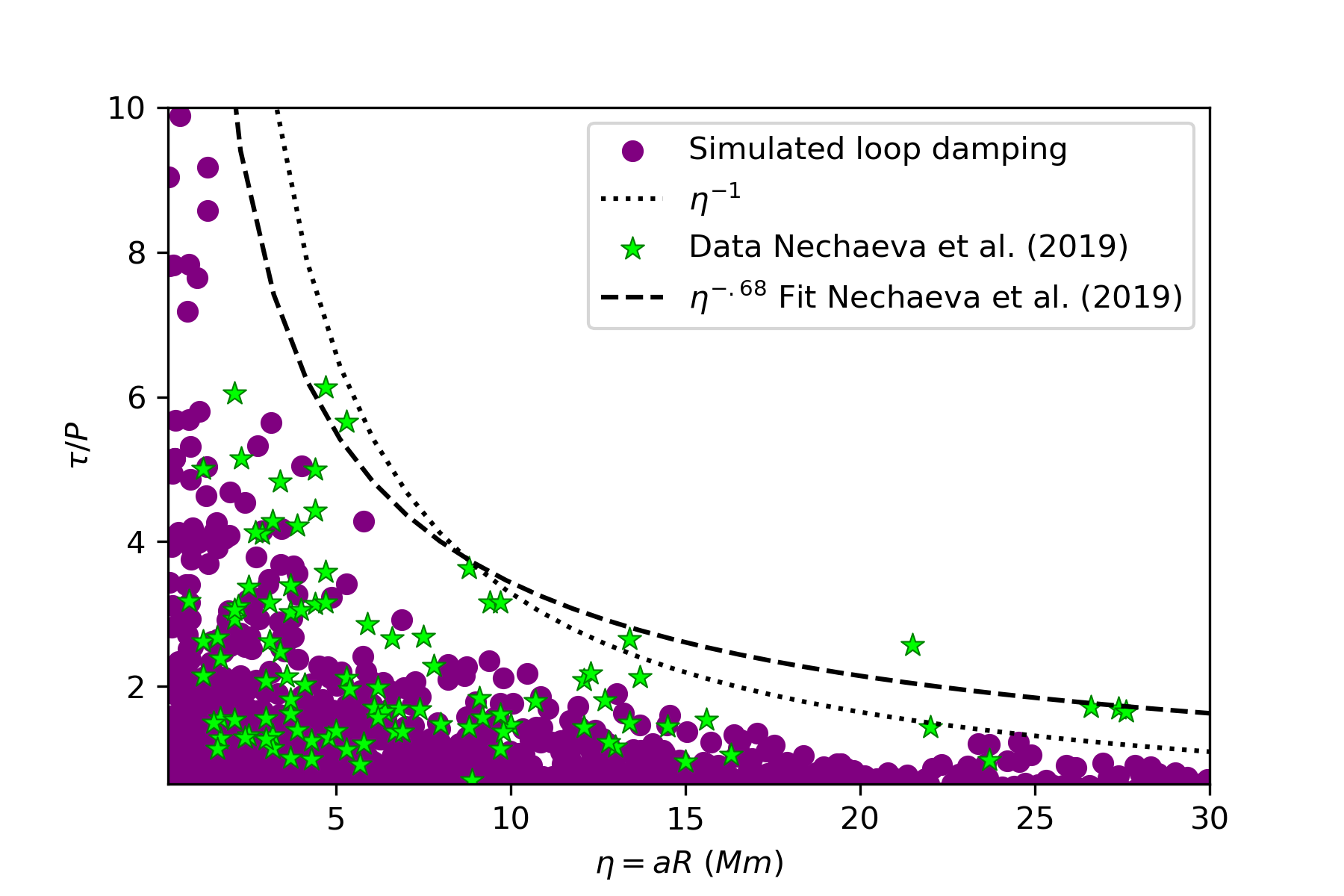}
	\caption{Scatter plot of the quality factor $\tau/P$ vs. the amplitude $\eta$ for the data of \citet{nechaeva2019} in green stars. Monte Carlo simulation of damping of kink waves in our current model with purple dots (left panel: only non-linear damping, right panel: harmonic average of non-linear damping and resonant absorption). The dotted line shows the possible outer envelope with $\eta^{-1}$ we obtained in this work and the dashed line shows the outer envelope $\eta^{-.68}$ as determined by \citet{nechaeva2019}.}
	\label{fig:forward}
\end{figure}

In this section, we investigate how the theory fits with data of damping of standing loop oscillations. We use the catalog mentioned in \citet{nechaeva2019}, and displayed as green stars in Fig.~\ref{fig:forward}. To generate the data points from our analytical theory, we follow the approach of \citet{verwichte2013}. They performed a Monte Carlo simulation for the resonant damping $\tau_\mathrm{RA}$ and period $P$ in which the loop parameters are drawn randomly from given distributions. Here we use the same procedure in order to study how $\tau/P$ depends on the amplitude $\eta$.\par 

We take 5000 realisations of oscillating coronal loops. We take as random variables for each loop: \begin{itemize}
	\item the density contrast $\zeta$ is drawn from a uniform distribution $[1,9.5]$, where the latter value for $\zeta_\mathrm{max}$ is taken from \citet{verwichte2013},
	\item the thickness of the inhomogeneous layer $l/R$ is drawn from a uniform distribution $[0,2]$,
	\item the amplitude $\eta$ is drawn from a uniform distribution between $[0.2,30]\rm{Mm}$, as suggested from the data,
	\item and the radius $R$ is uniformly drawn from a distribution between $[0.5,5]\rm{Mm}$.
\end{itemize}
All of these distributions are very crude. In principle, we could take the more advanced Bayesian inferences of \citet{pascoe2018}, but at this stage we have chosen to keep the results as simple as possible. Moreover, we have not taken into account projection effects on the oscillation amplitude, which would push down the horizontal scale by the cosine of the viewing angle. \\
For each randomly generated loop, we have plotted the non-linear damping time $\tau/P$ in the left panel of Fig.~\ref{fig:forward}, while the right panel shows the harmonic average of the non-linear damping time and resonant absorption, following the results of Sec.~\ref{sec:simulations}. \\
After drawing the random numbers and computing the associated damping times with the non-linear damping time or the harmonic average, we have added a 50\% noise to the damping rate $\tau/P$ to mimic observational constraints. With this, we mean that we multiply the theoretically obtained damping rate for the $i$th loop $(\tau/P)_i$ with $1+0.5n_i$, where the noise $n_i$ for the $i$th loop is drawn from the standard normal distribution $N(0,1)$. The specific shape of the noise (relative or absolute) does not have a lot of effect on our results.  
\par

The results of this Monte Carlo process are shown in Fig.~\ref{fig:forward}. There seems to be a good match between the green stars and purple dots in a statistical sense. The left panel shows an overpopulation towards long damping times, but this is solved in the right panel where the resonant absorption is taken into account. The lower part of the graph is densely populated with simulated data points, and this seems to lack in the observations. However, these points correspond to damping times $\tau/P<1$, which are hard to observe, and would rather be classified as non-oscillatory loops. Therefore, the observational points are probably biased towards the higher values of $\tau/P$. Consequently, the discrepancy between the purple points and the green points for small $\tau/P$ is a result of the observational bias. 

\citet{nechaeva2019} provided a fit for the outer bound of the data cloud, and found that it was given by $\eta^{-.68}$ (shown as a dashed black line in Fig.~\ref{fig:forward}). To contrast, we have overplotted the outer envelope with an $\eta^{-1}$ shape, as was found in our Eq.~\ref{eq:tau2}. Indeed, it seems that the $\eta^{-1}$ matches better with the purple simulated points in the left panel. However, moving to the right panel where resonant absorption is also taken into account, it is hard to say which power law fits better. Intuitively speaking, we could say that $\eta^{-.68}$ is an observational power law that results from averaging the $\eta^{-1}$ power law of non-linear damping and the $\eta^0$ power law of resonant absorption.\\
In any case, the data does not seem to contradict the current theory of non-linear damping of standing kink oscillations. However, Fig.~\ref{fig:appendix} for the alternative theory also provides a reasonable match of the data to the theoretical points. Perhaps a Bayesian model comparison approach \citep{montessolis2017} could confirm that the data fits better with our model (based on physical reasoning) than the model in the appendix, and this should be investigated in future work. 

\section{Conclusions}
In this paper, we have started from our previous results on the description of kink waves through \elsasser variables \citep{vd2020}. We have used those results to compute the energy density and non-linear energy cascade rate for standing kink waves. By taking the ratio of those, we have computed the predicted damping time of standing kink waves by the non-linear development of KHI. We found that the damping time is proportional to the period, and a complicated dependence on the density contrast, but that the damping does not vanish for density contrast equal to 1 (density of loop is the same as the exterior). More importantly, we found that the damping time is inversely proportional to the amplitude of the kink oscillation. \par

It was also found previously in simulations \citep{magyar2016b} and observations \citep{goddard2016, nechaeva2019} that the damping of kink waves gets stronger for increasing amplitude. In this paper, we have confronted our derived damping time with both these simulation and observational results. \\
From the comparison with the simulations of \citet{magyar2016b}, we found that the non-linear damping dominates for high amplitude oscillations, but that resonant absorption plays the dominant role for low amplitudes. Our analytical results capture the behaviour of the numerical results reasonably well, maybe aside from a vertical offset, which could be caused by a difference of the assumed transition layer. We established that taking the harmonic average of the resonant absorption damping time and the non-linear damping time gives a good match with the overall behaviour of the simulated loops. \\
We also compared our analytical formula to the results of \citet{nechaeva2019}. We have generated random loops in a Monte Carlo process, by varying the loop radius, loop density contrast, loop inhomogeneity, and oscillation amplitude. For these random loops, we have computed the expected non-linear damping time. From the comparison of the simulated loops with the observed loops, we see that there is a good match between the two data clouds. This includes the upper bound of the data cloud that is modelled in our case with $\eta^{-1}$, where $\eta$ is the amplitude of the oscillation.\par

Continuing on the Monte Carlo process for re-creating the observed data points, our method offers perspective on a Bayesian inference of loop parameters \citep[see e.g.][]{verwichte2013,arregui2013,pascoe2020} by finding the best statistical match between the two data clouds. This has the potential to seismologically determine the radii of coronal loops, the substructure of the loop, their density contrast distribution and the oscillation amplitude distribution. We will investigate the possibilities in a future work. \\
Further future work is in the numerical verification of the predicted non-linear damping rate of kink waves due to the formation of TWIKH rolls (Eq.~\ref{eq:tau2}). In the simulations of \citet{magyar2016b}, the effects of resonant damping and non-linear damping are intermixed, as was shown in Sec.~\ref{sec:simulations}. However, it is possible to run specialised simulations with $l/R\simeq 0$ to eliminate the influence of resonant absorption. We plan to run those simulations in the near future. 

\acknowledgments
TVD was supported by the European Research Council (ERC) under the European Union's Horizon 2020 research and innovation programme (grant agreement No 724326) and the C1 grant TRACEspace of Internal Funds KU Leuven. The research benefitted greatly from discussions at ISSI-BJ.


%\vspace{5mm}
%\facilities{HST(STIS), Swift(XRT and UVOT), AAVSO, CTIO:1.3m,


%\software{astropy \citep{2013A&A...558A..33A},  
%          Cloudy \citep{2013RMxAA..49..137F}, 
%          SExtractor \citep{1996A&AS..117..393B}
%          }



\appendix

\section{Non-linear damping of standing kink waves: alternative calculations}\label{sec:appendix}

In Sec.~\ref{sec:damping}, we have explained that there is no mathematical preference in the averaging and summing of the two \elsasser components. We have taken the physically correct approach to first add the contributions before averaging them and these are the results we presented in the main text. This is because the kink wave contains both \elsasser components, which are damped by both $\epsilon^\pm$ simultaneously. For completeness, we also list the alternative computations even though we think that these are incorrect.  \\
Mathematically, the problem is the non-commutativity of the averaging and summing. We can write down that 
\begin{equation}
	\langle \langle \epsilon \rangle \rangle = \langle\langle \epsilon^+ +\epsilon^-\rangle\rangle \neq \langle\langle \epsilon^+ \rangle\rangle +\langle\langle \epsilon^-\rangle\rangle.
\end{equation}
In the main text of this paper, we have taken the left hand side of the inequality sign. Here we consider the right hand side. 

As before, we start from the result in Eq.~\ref{eq:energycascade}. We have subsequently:
\begin{align}
	\langle\langle \epsilon^\mp \rangle\rangle 
	& = V^3 \frac{\sqrt{\pi} R}{20} \frac{\rho_\mathrm{e}}{\omega^3} \left(\frac{k_z}{2\pi}\int dz (\omega^2 \cos^2{k_z z}+\omega_\mathrm{Ae}^2 \sin^2{k_z z})( 5(\omega^2 \cos^2{k_z z}+ \omega_\mathrm{Ae}^2 \sin^2{k_z z})^2-16 \omega^2\omega_\mathrm{Ae}^2 \cos^2{k_z z} \sin^2{k_z z})\right)^{1/2}\\
	&= V^3 \frac{\sqrt{\pi} R}{20} \frac{\rho_\mathrm{e}}{32}\sqrt{25\zeta^3+73\zeta^2+67\zeta+219}
\end{align}
As in the main text, the damping does not go to 0 as $\zeta\to 1$, also confirming in this case the results of \citet{howson2019} that the non-linear damping by KHI also works in a uniform density. \par

\begin{figure}
	\plotone{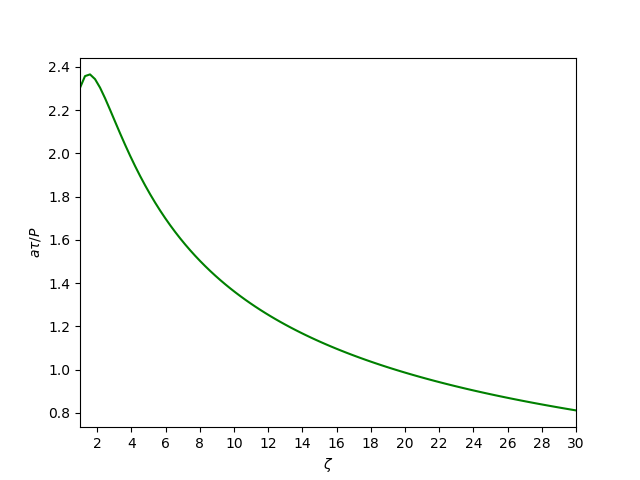}
	\caption{$a\tau /P$ for a standing kink wave as a function of the density contrast $\zeta$, following formula Eq.~\ref{eq:tau1}.}
	\label{fig:tau1}
\end{figure}

\begin{figure}
	\plotone{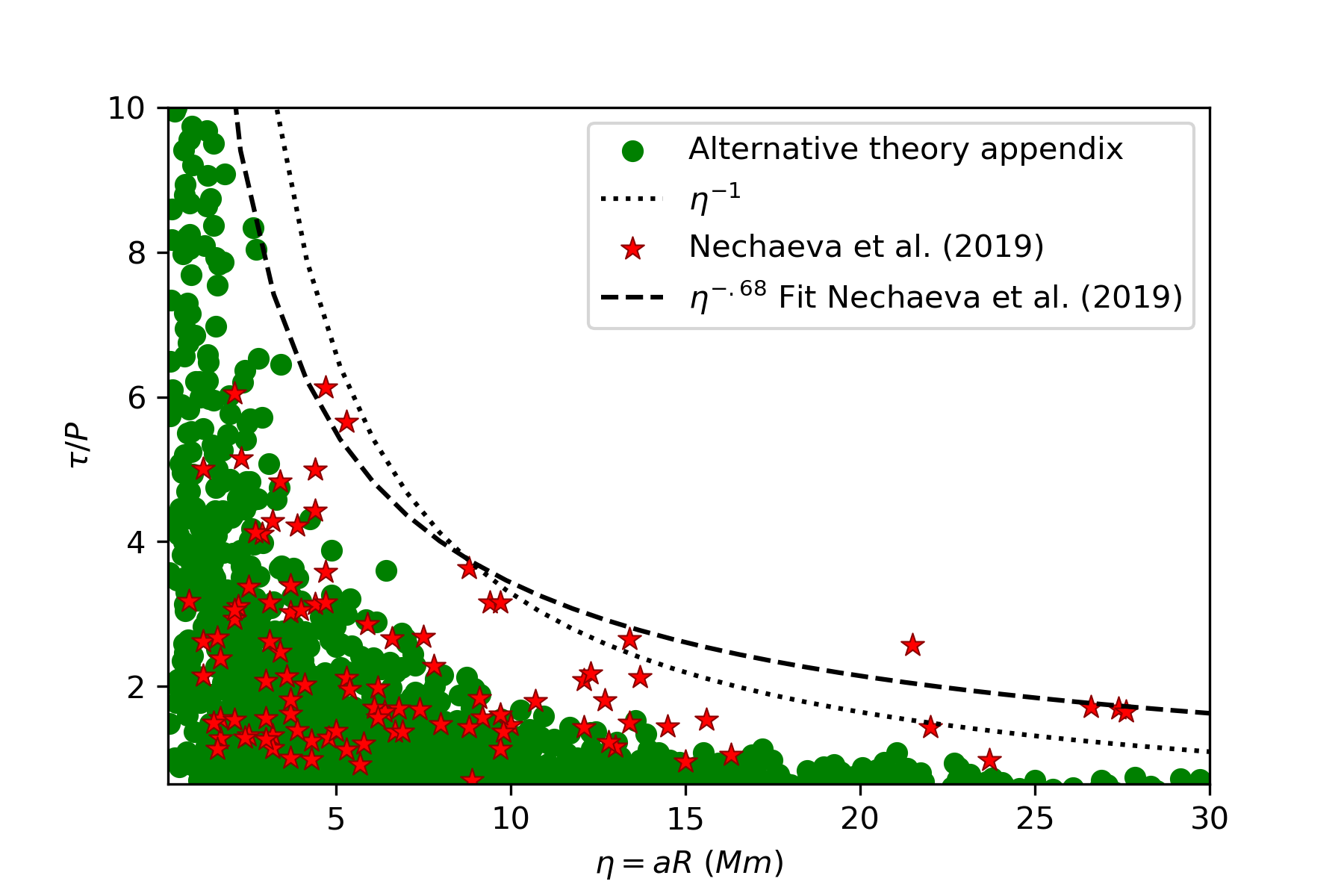}
	\caption{Scatter plot of the quality factor $\tau/P$ vs. the amplitude $\eta$ for the data of \citet{nechaeva2019} in red stars. Monte Carlo simulation of damping of kink waves in the appendix with green dots. The dashed and dotted lines are the same as in Fig.~\ref{fig:forward}.}
	\label{fig:appendix}
\end{figure}

Now we can find the expression for the damping time by dividing the energy by the energy cascade rate, using the same approach as before:
\begin{equation}
	\tau=\frac{\langle w\rangle}{\langle\langle \epsilon\rangle\rangle}=80\sqrt{\pi}\frac{R}{V}\frac{1+\zeta}{\sqrt{25\zeta^3+73\zeta^2+67\zeta+219}}= 80\sqrt{\pi}\frac{P}{2\pi a}\frac{1+\zeta}{\sqrt{25\zeta^3+73\zeta^2+67\zeta+219}}.\label{eq:tau1}
\end{equation}
Remarkably, the dependence on the density contrast is entirely different than before. It is shown graphically in Fig.~\ref{fig:tau1}.
The comparison with the simulated data is shown with the thin green lines in Fig.~\ref{fig:magyar}, showing a somewhat better match with the data. For completeness, we also show in Fig.~\ref{fig:appendix} the simulated data points with the Monte Carlo method of Sec.~\ref{sec:monte}. The fit of the simulated data with the observed data points is somewhat less than in Fig.~\ref{fig:forward}, but not enough to distinguish the two theories, leaving it to physical grounds alone. Still, we believe this alternative theory in the appendix to be incorrect. 


\bibliography{../../refs/refs}
\bibliographystyle{aasjournal}



\end{document}